# How to infer ocean freezing rates on icy satellites from measurements of ice thickness

Nicole C. Shibley,[1,2,3]★ Ching-Yao Lai,[4,5] and Riley Culberg[4,6]
[1] Princeton Center for Theoretical Science, Princeton University, Princeton, NJ 08544, USA
[2] (currently) Department of Applied Mathematics & Theoretical Physics, University of Cambridge, Cambridge, CB3 0WA, UK
[3] (currently) Department of Earth Sciences, University of Cambridge, Cambridge, CB2 3EQ, UK
[4] Department of Geosciences, Princeton University, Princeton, NJ 08544, USA
[5] (currently) Department of Geophysics, Stanford University, Stanford, CA 94305, USA
[6] (currently) Department of Earth and Atmospheric Sciences, Cornell University, Ithaca, NY 14853, USA

Accepted XXX. Received YYY; in original form ZZZ

## ABSTRACT

Liquid-water oceans likely underlie the ice shells of Europa and Enceladus, but ocean properties are challenging to measure due to the overlying ice. Here, we consider gravity-driven flow of the ice shells of icy satellites and relate this to ocean freeze and melt rates. We employ a first-principles approach applicable to conductive ice shells in a Cartesian geometry. We derive a scaling law under which ocean freeze/melt rates can be estimated from shell-thickness measurements. Under a steady-state assumption, ocean freeze/melt rates can be inferred from measurements of ice thickness, given a basal viscosity. Depending on a characteristic thickness scale and basal viscosity, characteristic freeze/melt rates range from around O($10^{-1}$) to O($10^{-5}$) mm/year. Our scaling is validated with ice-penetrating radar measurements of ice thickness and modelled snow accumulation for Roosevelt Island, Antarctica. Our model, coupled with observations of shell thickness, could help estimate the magnitudes of ocean freeze/melt rates on icy satellites.

## 1 INTRODUCTION

Several icy satellites exist in the solar system. Europa and Enceladus, in particular, have generated significant interest due to their young ice covers thought to be overlying liquid-water oceans (e.g., Cassen et al. 1979; Carr et al. 1998; Pappalardo et al. 1999; Kivelson et al. 2000; Porco et al. 2006; Postberg et al. 2009; Roth et al. 2014). Such speculation has prompted interest in these satellites as possible locations for extraterrestrial life (e.g., Hand et al. 2009; Cable et al. 2021). However, there are significant first-order questions which have yet to be answered about both the ice shells and oceans of these satellites, which may help constrain future questions about astrobiology.

One key question is the thickness of the satellites' ice shells and whether or not this thickness varies spatially. On Europa, generally, the ice shell is thought to be between ~3 km (e.g., Hoppa et al. 1999; Schenk 2002) and ~ 30 km thick (e.g., Ojakangas & Stevenson 1989; Schenk 2002; Pappalardo et al. 1998; Howell 2021). Due to the lower surface temperature at the pole than at the equator, it is expected that the ice shell may be thicker near the poles than near the equator (Ojakangas & Stevenson 1989); this gradient in ice thickness may result in spatially-varying ocean stratification (Zhu et al. 2017). The presence of a lateral ice thickness gradient is also thought to occur on Enceladus (e.g., Hemingway & Mittal 2019; Beuthe 2018), with modelling results based on inferences from observations suggesting pole-to-equator thickness differences of between 5 to 30 km (Hemingway & Mittal 2019). The ability of the ocean to transport heat meridionally can ultimately homogenize the ice thickness in either scenario (Kang & Jansen 2022).

When ice exhibits horizontal gradients in thickness, on long enough timescales, it can flow as a viscous fluid (Pegler & Worster 2012; Worster 2014), similar to how syrup spreads on a pancake. This is due to a gravity-driven flow from regions of high pressure (thick ice) to regions of low pressure (thin ice), known as a gravity current (e.g., Huppert 1982). Gravity currents are ubiquitous in nature and describe many natural phenomena ranging from cold fronts (Simpson & Britter 1980), to mantle intrusions (Kerr & Lister 1987) to glacial flow (Kowal & Worster 2015).

In the context of icy satellites, several past studies have considered how ice flow may be invoked to understand surface topography (e.g., Stevenson 2000; Nimmo 2004; Nimmo & Bills 2010; Čadek et al. 2019) and the underlying ocean (Kamata & Nimmo 2017; Ashkenazy et al. 2018; Čadek et al. 2019; Kang et al. 2022). In particular, the two-dimensional and three-dimensional general circulation modelling studies of Ashkenazy et al. (2018); Kang & Jansen (2022); Kang (2022); Kang et al. (2022) have related the lateral ice flow on icy satellites to ocean dynamics (and vice versa), considering cases with both meridional ocean heat transport and ice convection (Ashkenazy et al. 2018), tidal heating in the shell (e.g., Ashkenazy et al. 2018; Kang et al. 2022), the effects of gravity (Kang & Jansen 2022), and oceanic eddy transport (Kang 2022). Such general cir-culation models have the advantage of simulating multiple physical processes of a complex system in a global setup. However, a limitation of such models is the number of free parameters inherent to the system, making it challenging to invert for any single parameter given a set of observations.

★ E-mail: nicole.shibley@damtp.cam.ac.uk

More specifically, the modeling study of Ashkenazy et al. (2018) consider a conductive (along with convective) spherical setup for Europa's ice shell. In their seminal work, they describe how Europan ice thicknesses would look under various ice-ocean configurations (for example, a conductive versus convective ice shell, with varying effects of internal heating, with/without ocean heat transport, and with different ocean diffusivities), incorporating the effects of ocean freezing/melting into their model equations. In our study, we consider the converse: if one *knows* the ice-thickness distribution from observations, what can realistically be said about the ocean, specifically about freeze and melt rates?

Thus, here we attempt to distill the governing physics for a purely-conductive shell with a temperature-dependent viscosity to provide an understanding of the simplest ocean parameters which can realistically be inferred from future, expected ice-thickness measurements. A two-dimensional floating viscous gravity current is considered, where ice flows from pole to equator and under which an ice thickness gradient can be sustained by spatially-varying freezing and melting. The value of our O(0) simplified model is that it requires only one choice of free parameter (the basal viscosity, $\eta_b$), while containing all information about the ocean into a freezing and melting term, $b(x)$. We examine how freeze and melt rates can be inferred from lateral thickness gradients in a steady-state, and relate these to different viscosity regimes. We further explicate a scaling law which describes how freeze/melt rates can be estimated from ice thickness scales. Finally, our simplified model and scaling are compared to Earth-based radar observations of Antarctica to corroborate our results.

## A SIMPLIFIED ICE-OCEAN MODEL

We consider a simplified setup, with an "inviscid" ocean underlying the ice shell. The ice shell experiences a temperature gradient across it since the surface temperature is much colder than the basal temperature (e.g., Ashkenazy 2019), leading to a depth(temperature)-dependent viscosity (e.g., Goldsby & Kohlstedt 2001). This leads to an upper, brittle ice lid under which sits a flowing, viscous ice layer (Figure 1).

To illustrate the dynamics, we consider a two-dimensional setup, where ice flows laterally from pole to equator (Figure 1). This setup is predicated on the assumption that there is thick ice at the pole and thin ice at the equator, as in Ojakangas & Stevenson (1989), for example; such a gradient would be set up by a pole-to-equator temperature difference with colder temperatures at the pole and warmer temperatures at the equator (see e.g., Ashkenazy et al. 2018; Ashkenazy 2019). We note that studies which have considered the influence of ice convection have suggested that it may be possible to setup the reverse gradient, with thicker ice at the equator and thinner ice at the pole (Ashkenazy et al. 2018). Our analysis applies only to a conductive system. Thus, we restrict our values of basal viscosity and thickness to values which can generally be expected to be in a conductive heat-transport regime (see e.g., Shibley & Goodman 2024; McKinnon 1999). We further note that we do not consider an ice-pump mechanism (Lewis & Perkin 1986) here.

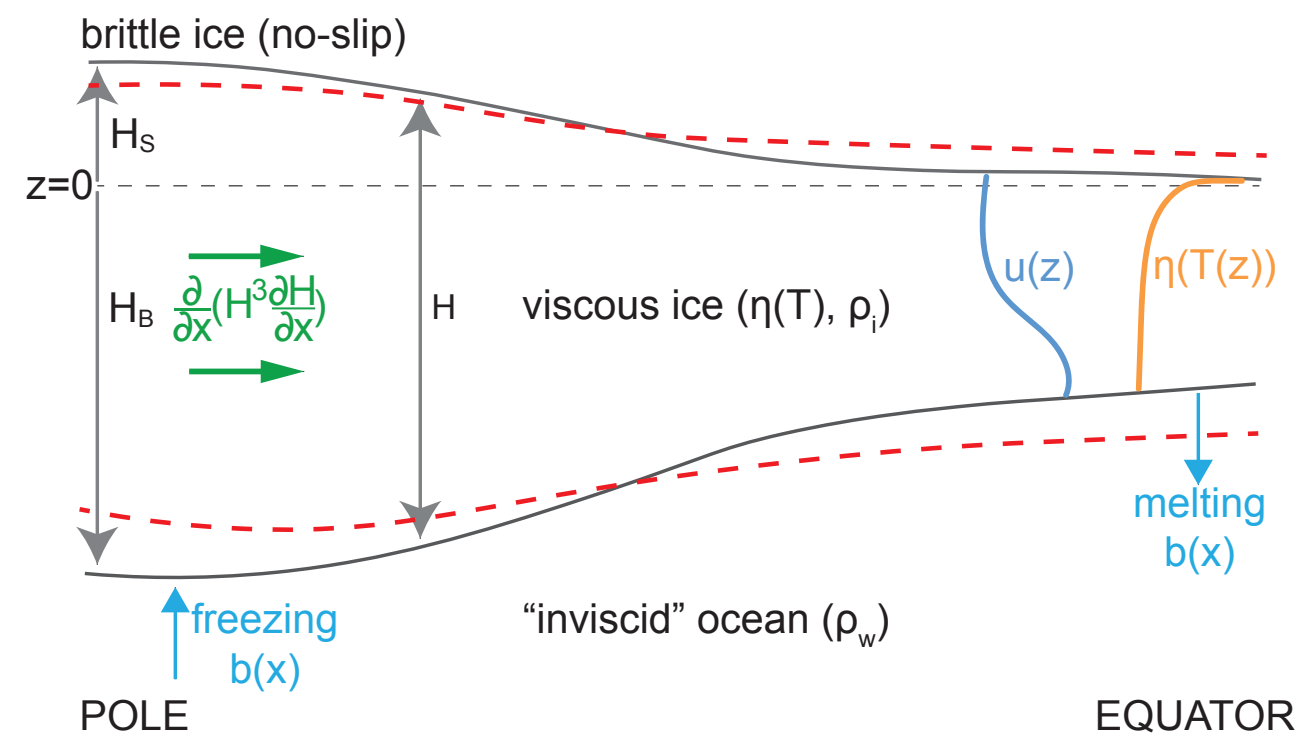

**Figure 1.** Schematic of ice shell flow in a 2D setup. The pole is shown on the lefthand side and equator on the righthand side of the plot. The ice thickness above sea level (at $z = 0$) is $H_S$ and below sea level is $H_B$. $H = H_S + H_B$ is the total ice thickness. Viscous ice, with viscosity $\eta(T)$ and density $\rho_i$, sits atop an "inviscid" ocean of density $\rho_w$. The viscous ice shell either thins as it spreads or is sustained by freezing at the pole and melting at the equator. A new shell thickness, governed by the green nonlinear diffusive term, is shown by the dashed red line. A spatially-varying freeze/melt function is given by $b(x)$. A schematic representation of the viscosity profile $\eta(T)$ is shown in the orange line, and the resultant horizontal velocity profile $u(z)$ in the blue line.

### Mathematical Formulation

The system can be described by the following equations, which generally follow the standard gravity current equations of Huppert (1982) and which we extend to include the effect of a temperature-dependent viscosity. The total thickness of the ice shell is $H = H_S + H_B$, where $H_S$ falls above the $z = 0$ sea level, and $H_B$ falls below the $z = 0$ line. Then, $H_S = (1 - \rho_i/\rho_w)H$, where $\rho_i = 920$ kg m$^{-3}$ is the density of ice, and $\rho_w = 1000$ kg m$^{-3}$ is the density of the ocean.

We start with conservation of mass: $\frac{\partial u}{\partial x} + \frac{\partial w}{\partial z} = 0$, and conservation of momentum for Stokes flow, combined with the fact that the horizontal length scale is much larger than the vertical length scale, leading to:

$$0 = -\frac{\partial p}{\partial x} + \frac{\partial}{\partial z}(\eta(T(z))\frac{\partial u}{\partial z}), \text{ and } 0 = -\frac{\partial p}{\partial z} - \rho_i g\hat{z}, \quad (1)$$

where $p$ is pressure, $u$ is the velocity in the $x$-direction, $w$ is the velocity in the $z$-direction, $\eta(T)$ is a temperature-dependent viscosity, $\rho_i$ is ice density, and $g$ is gravity. We present our method in Cartesian coordinates here for ease of understanding, but the same physics holds regardless of coordinate system.

A temperature-dependent (or equivalently depth-dependent) viscosity is appropriate since the upper surface of the ice shell will be significantly colder than the base. Here we employ the Frank-Kamenetskii approximation (e.g., Jain & Solomatov 2022), defined as:

$$\eta(T) = \eta_b e^{(1/l)(1-\frac{T-T_S}{\Delta T})}, \quad (2)$$

where $\eta_b$ is the specified basal viscosity, $\tilde{T} = \frac{T-T_S}{\Delta T}$, $\Delta T = T_B - T_S$, $T_B = 273$ $K$ is the ice temperature at the ice-ocean interface, $T_S = 93$ $K$ is an ice surface temperature (appropriate for Europa, Ashkenazy 2019), and $l = \frac{RT_B^2}{Q\Delta T}$, where $R$ is the gas constant and $Q = 60$ kJ mol$^{-1}$ is the activation energy. For Enceladus, an appropriate ice surface temperature would be $\sim 60$ K (e.g., Hemingway & Mittal 2019). We note here that we take $\Delta T$ to be constant across the entire ice shell, though lateral variations in the temperature jump will exist; these are what give rise to variations in ice thickness in the first place. This approximation allows for the equation to be solved analytically. We note, however, that if we consider the case where the change in temperature jump is ∼223 K (i.e., taking the surface temperature to be a Europan polar surface temperature, 50 K) and which would really only be relevant near the pole, the main results do not vary significantly. The freeze-melt scaling law discussed later returns slightly lower values in the polar surface temperature case, but approximately the same order of magnitude regardless. This is likely because the effect of the increased effective viscosity (due to the larger temperature jump) is most pronounced closest to the surface, where the ice shell is already not dynamically very interesting.

The relevant boundary conditions are no-slip at the top surface (brittle lid), and free-slip (no-stress) at the ocean-ice interface, given by: $u(z = H_S) = 0$, and $\frac{\partial u}{\partial z}|_{z=-H_B} = 0$.

Then, taking $p = p_o + \rho_i g(H_s - z)$, where $p_0$ is a reference pressure, yields:

$$0 = -\rho_i g \frac{\partial H_S}{\partial x} + \frac{\partial}{\partial z}(\eta_b e^{[\frac{1}{l}(1-\frac{H_S-z}{H})]}\frac{\partial u}{\partial z}), \tag{3}$$

where $\frac{H_S-z}{H} = \frac{T-T_S}{\Delta T}$ for a linear temperature gradient across the shell.

Then, nondimensionalizing where $\tilde{z} = \frac{(H_s-z)}{H}$, $d\tilde{z} = -\frac{dz}{H}$, and $\tilde{u} = \frac{-u\eta_b e^{1/l}}{\rho_i g H^2(\partial H_S/\partial x)}$ gives:

$$0 = 1 + \frac{\partial}{\partial \tilde{z}}(e^{-\tilde{z}/l}\frac{\partial \tilde{u}}{\partial \tilde{z}}), \tag{4}$$

with boundary conditions: $\frac{\partial \tilde{u}}{\partial \tilde{z}}|_{\tilde{z}=1} = 0$, and $\tilde{u}(\tilde{z} = 0) = 0$. This can be solved to give an analytical dimensionless velocity profile:

$$\tilde{u} = l[e^{\tilde{z}/l}(l + 1 - \tilde{z}) - (l + 1)] \tag{5}$$

Finally, depth-integrating the mass conservation equation gives:

$$\frac{\partial q}{\partial x} + \frac{\partial H}{\partial t} = b(x), \tag{6}$$

where $q = \int_{-H_B}^{H_S} u dz$ and $b(x)$ is the source/sink term that describes background freeze and melt rates from the ocean.

This can be rewritten as:

$$\frac{\partial H}{\partial t} - \frac{\rho_i g}{\eta_b e^{1/l}}\beta\frac{\partial}{\partial x}[H^3\frac{\partial H_S}{\partial x}] - b(x) = 0, \tag{7}$$

where $\beta = \int_0^1 \tilde{u}d\tilde{z} = l[2l^2(e^{1/l} - 1) - 2l - 1]$.

Finally, defining $\gamma = \beta e^{-1/l}$ and recalling that $H_S = (1-\rho_i/\rho_w)H$ (note this balance assumes isostasy, relevant to large-scale/long-wavelength thickness changes, see e.g., Nimmo et al. (2007)), we obtain the following:

$$\frac{\partial H}{\partial t} - \frac{\rho_i g\gamma}{\eta_b}(1 - \frac{\rho_i}{\rho_w})\frac{\partial}{\partial x}[H^3\frac{\partial H}{\partial x}] - b(x) = 0. \tag{8}$$

This equation describes how the thickness of an ice shell with a temperature-dependent viscosity changes in time as it flows laterally and is modified by spatially-varying freezing and melting.

## RESULTS

Changes in shell thickness due to gravity-driven flattening compete with thickening and thinning of the shell due to freezing and melting of ice driven by ocean heat fluxes (see also Kang & Jansen 2022; Kang 2022). Given the temperature-dependence of viscosity and the temperature gradient across the shell, the flowing portion of the shell is confined to the bottom of the shell (Figure 2a); this can also be seen from the functional form of Equation 5. The flow rate varies in space and in time depending on the evolving local thickness gradient. It is also larger for lower basal viscosities, and smaller for higher basal viscosities. For example, for the ice shell shown here after 100 million years with a basal viscosity of $10^{14}$ Pa s, flow rates vary from approximately 0 to 2.6 mm/year (Figure 2a).

Unless sustained by ocean freezing and melting, the ice thickness will homogenize over time (Figure 2b); ice shells with lower basal viscosities flatten faster than those with larger viscosities (Figure 2c). Recent work (Kihoulou et al. 2023) has also shown how larger effective viscosities can lead to steeper ice shell topography.

### Spatial Variation of Freeze and Melt Rates

Freezing and melting concurrent with ice flow modifies the shell-thickness distribution (Figure 3a). If ice freezes at the pole and melts at the equator at exactly the same rate as ice moves laterally between the two regions, then there will be no temporal change in ice thickness, and the ice will remain in steady-state. If an ice cover can be considered to be in steady state (a common assumption for planetary ice shells, e.g., Ojakangas & Stevenson 1989; Hussmann et al. 2002; Tobie et al. 2003; Ashkenazy et al. 2018; Hemingway & Mittal 2019; Akiba et al. 2022; Kang & Jansen 2022; Kang 2022), this then provides a method by which to calculate both the spatial distribution of ocean freeze and melt rates based on global measurements of ice thickness, as well as a way to infer the magnitudes of maximum ocean freeze/melt rates and their spatial locations.

Ocean freezing or melting results when the ice-ocean heat flux at the ice-ocean interface and the vertical conductive flux out of the base of ice shell are out of balance (see e.g., Shibley & Goodman 2024, for a simplified vertical model). The magnitude of ice-ocean heat flux arises from ocean heat-transport processes (such as turbulent mixing or molecular diffusion) which redistribute heat from the ocean towards the ice-ocean interface. Several studies incorporate ocean mixing parametrizations into their modeling of planetary ice or ocean dynamics (Kang 2022, 2023; Zeng & Jansen 2024). Here, we instead *infer* ocean freeze and melt rates, which result from ocean mixing/heat-transport processes, *from* the ice thickness measurements. This term $b(x)$ captures the effect of ocean heat transport, which causes the freezing and melting, without requiring a need for a particular knowledge of specific ocean processes. In fact, these inferred freezing and melting rates could offer insight or validation of hypothesized subsurface mixing processes in the ocean. We describe how steady-state measurements of ice thickness can be used to infer information about ocean freezing and melting next.

The freeze/melt rate necessary to result in a steady-state ice thickness can be calculated via the following equation (see Equation 8, Figure 3b):

$$b(x) = -\frac{\rho_i g\gamma}{\eta_b}(1 - \frac{\rho_i}{\rho_w})\frac{d}{dx}[H^3\frac{dH}{dx}]. \tag{9}$$

The spatial location of the maximum or minimum freeze/melt rate can then be found by solving:

$$\frac{d}{dx}[H^3\frac{d^2H}{dx^2} + 3H^2(\frac{dH}{dx})^2] = 0. \tag{10}$$

The magnitude of the freeze and melt rate can then be determined via Equation 9. This means that a sufficiently-resolved global ice-thickness distribution $H(x)$ would indicate the locations $(x)$ of maximum/minimum melt rates in an icy satellite ocean.

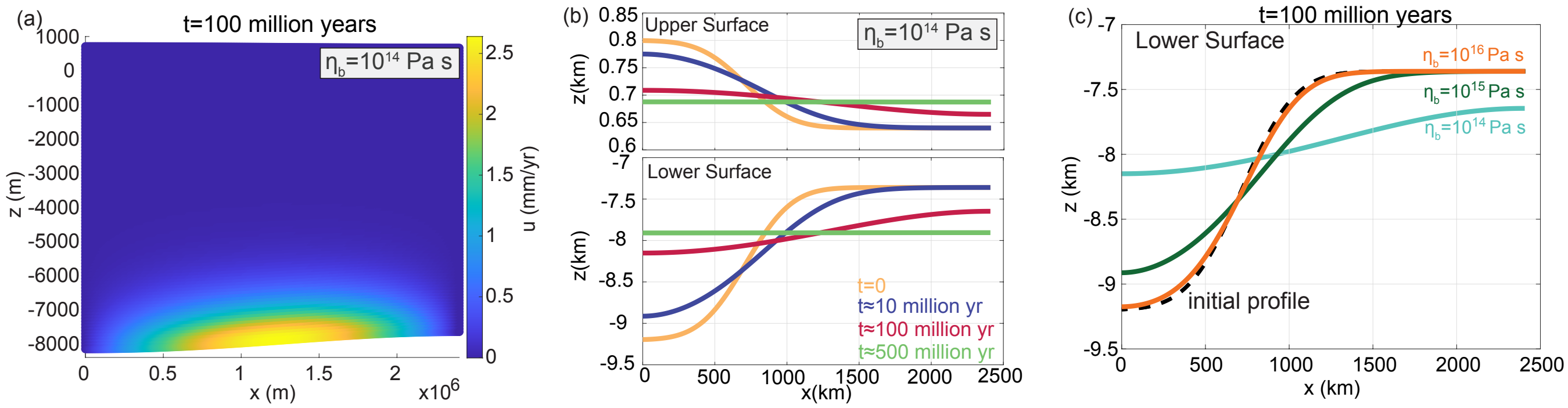

**Figure 2.** (a) Flow field after 100 million years for an ice shell with a temperature-dependent viscosity, with basal viscosity of $\eta_b = 10^{14}$ Pa s. Since the bottom of the shell is warmer than the upper surface, the base is less viscous, and the flow is concentrated in this portion of the shell. For the specified basal viscosity at time $t = 100$ million years, the maximum flow rate is approximately 2.6 mm/year. (b) Surface and basal topography of an ice shell, with an initial topography shown at $t = 0$ (yellow), for $\eta_b = 10^{14}$ Pa s at different times (colors). The ice shell flattens over time. (c) Basal topography of an ice shell, with an initial topography shown in (b), at different values of basal viscosity $\eta_b$ after 100 million years. While shells with lower viscosities have flattened, shells with higher viscosities still maintain a similar profile to the initial state. The example here is based on parameters appropriate for a Europan ice shell.

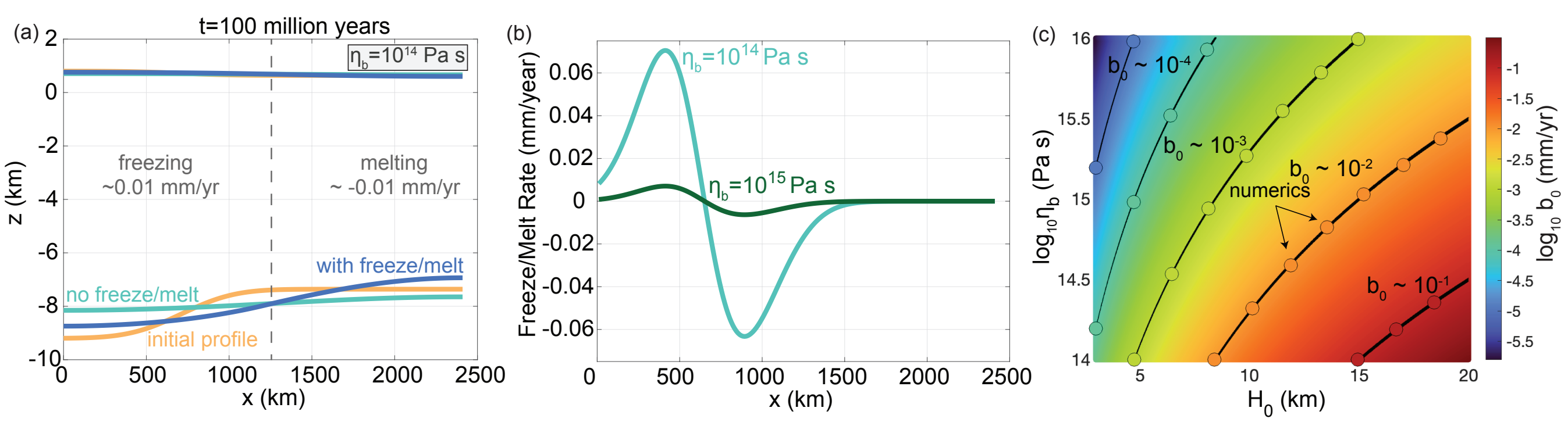

**Figure 3.** (a) Flow of an ice shell, with an initial thickness shown at $t = 0$ (yellow), for $\eta_b = 10^{14}$ Pa s after 100 million years both with the presence of freezing/melting (blue) and without (aqua). The ice shell flattens over time in the absence of freezing and melting, but maintains a thickness gradient when ocean freeze/melt are included. (b) Freeze and melt rate needed to maintain a steady-state thickness gradient (shown by the yellow profile in (a)). Ice shells with lower basal viscosities must have larger freeze/melt rates to offset the flow of ice. (c) Scaling for freeze/melt rate (color, mm yr$^{-1}$) of ice shell flow for different thickness scales and basal viscosities. Black lines show contours of $b_0 \sim \frac{\gamma\rho_i g}{\eta_b}(\frac{\rho_w-\rho_i}{\rho_w})\frac{H_0^4}{L_0^2}$, and colored dots show the calculated freeze/melt estimate, $\overline{|b(x)|}$, plotted on top. A linear fit between the calculated estimate and theoretical scaling indicates a prefactor $k$ of 1.1, for the particular profile $H(x)$ used here. The example here is based on parameters appropriate for a Europan ice shell.

**Scalings of Ocean Dynamics**

Further, how the freeze and melt rate depends on both ice shell basal viscosity, as well as ice shell thickness, can be described by a scaling law. Such a law arises by considering characteristic scales for each of the terms in Equation 8 and balancing them against each other; the scaling has the advantage of necessitating sparse ice-thickness measurements to make estimates of the freeze/melt rate and does not require a steady-state assumption. This means that even if upcoming space missions do not return a robust spatial map of observations, a scaling may still be used to estimate freeze/melt rates on an icy satellite, provided that $\partial H/\partial t$ is not much larger than the other terms in Equation 8. Disadvantages of a scaling approach are that this does not give precise location measurements as to freezing and melting, and rather a single order-of-magnitude estimate, which furthermore is dependent upon knowledge of the basal viscosity (a parameter currently unknown for icy satellites).

A characteristic freeze and melt rate follows:

$$b_0 \sim \frac{\gamma\rho_i g}{\eta_b}\left(\frac{\rho_w-\rho_i}{\rho_w}\right)\frac{H_0^4}{L_0^2}, \tag{11}$$

where $b_0$ is a characteristic freeze-melt scale. Here we take a characteristic horizontal scale $L_0$ to be the pole-to-equator distance and a characteristic vertical scale $H_0$ to be the thickness of ice at the pole (i.e., the maximum ice thickness at any space or time). Further, recall that $\gamma = le^{-1/l}[2l^2(e^{1/l}-1)-2l-1]$, where $l = (RT_B^2)/(Q\Delta T)$. If a scaling prefactor $k$, whose value depends on the shape of $H(x)$, is included, this gives $b_0 = k\frac{\gamma\rho_i g}{\eta_b}(\frac{\rho_w-\rho_i}{\rho_w})\frac{H_0^4}{L_0^2}$. For the particular thickness profile examined here, we find $k = 1.1$, indicating a good fit between the scaling law and the estimated freeze and melt rate. This is taken from a linear fit between $\frac{\gamma\rho_i g}{\eta_b}(\frac{\rho_w-\rho_i}{\rho_w})\frac{H_0^4}{L_0^2}$ and the mean of $|b(x)|$ calculated from Equation 9, using the initial thickness profile (described in the Materials and Methods section) shown in Figure 3a. Note that the characteristic vertical scale $H_0$ is taken to be a vertical thickness rather than a thickness difference as for the profiles tested this yielded a calculated prefactor $k$ closer to 1. Although we are considering a simplified Cartesian problem here, the extension to a spherical coordinate system can be done without affecting the scaling in Equation 11. The exact magnitude of freeze and melt depends on the shape of the thickness profile.

In spherical coordinates, following the derivation described in Ashkenazy et al. (2018) assuming zonal symmetry, integrating from the base to the surface (instead of to a transition temperature, as in their work) and taking a relationship between stress $\tau$ and strain rate $\dot{\epsilon}$ as $\tau_{ij} = 2\eta(T)\dot{\epsilon}_{ij}$ (instead of $\tau_{ij} = \eta(T)\dot{\epsilon}_{ij}$), it can be shown that the governing equation in the spherical setup is:

$$\frac{\partial H}{\partial t} - \frac{\rho_i g \gamma}{\eta_b}(1 - \frac{\rho_i}{\rho_w})\frac{1}{R^2 sin\theta}\frac{\partial}{\partial \theta}[H^3 sin\theta \frac{\partial H}{\partial \theta}] - b(\theta) = 0, \quad (12)$$

where $\theta$ is co-latitude. Note the similarities between this equation and equation 8 in the Cartesian system. This gives a characteristic scale for melt/freeze rate of $b_0 \sim \frac{\gamma \rho_i g}{\eta_b}\frac{\rho_w - \rho_i}{\rho_w}\frac{H_0^4}{R_0^2}$, the same functional form as the Cartesian scaling, where now the length scale $R_0$ is taken to be the radius of the satellite. Note that these scalings apply to a system approximated as two-dimensional, where the shell thickness varies in one of the dimensions. If a scaling prefactor $k$ is included as in $b_0 = k\frac{\gamma \rho_i g}{\eta_b}\frac{\rho_w - \rho_i}{\rho_w}\frac{H_0^4}{R_0^2}$, it will be a scaling prefactor particular to the spherical setup (any coordinate system will have its own prefactor). We choose to present our methodology in a Cartesian basis for ease of explanation.

The scaling (Equation 11) indicates that with a characteristic shell thickness scale and with a knowledge of ice viscosity, a characteristic freeze/melt rate can be inferred for the ocean. For a larger characteristic vertical thickness $H_0$, the characteristic freeze and melt rate $b_0$ will be larger than for smaller vertical thicknesses at the same viscosity (Figure 3c); this is because shells for which the value $|3H^2(\frac{\partial H}{\partial x})^2 + H^3\frac{\partial^2 H}{\partial x}|$ is larger accumulate/flatten ice faster than at lower values.

However, the ability to infer the correct magnitude of ice shell freezing and melting is dependent on a knowledge of the ice viscosity which controls how quickly the ice spreads; this is not well-prescribed (see e.g., Shibley & Goodman 2024). For the same initial thickness profile, a shell with a lower basal viscosity will require a larger freeze/melt rate in order to sustain a steady-state shell thickness than a shell at larger viscosity (Figure 3b). For example, for the profile shown in Figure 3b, the maximum freeze/melt rate necessary to maintain a steady-state thickness is about 0.07 mm/year for a shell with $\eta_b = 10^{14}$ Pa s, whereas for a shell with basal viscosity $\eta_b = 10^{15}$ Pa s, the magnitude of the maximum freeze/melt rate necessary to maintain a steady-state is one order of magnitude smaller (about 0.007 mm/year). This is because the viscous portion of the shell will flow more slowly in the case of larger basal viscosity; thus, the rate of freezing and melting needed to offset this flow is smaller than for shells with a lower basal viscosity.

It is important to note that the ice shell viscosity does not physically determine the freeze and melt rate of the ice shell (this is governed by ocean dynamics, e.g., Ashkenazy et al. 2018; Kang 2022), but rather that a knowledge of the ice rheology is required in order to infer an ocean freeze/melt rate based on measurements of ice thickness. Depending on the basal ice shell viscosity, varying here from $10^{14}$ to $10^{16}$ Pa s and the characteristic vertical thickness scale (which will depend on the shell profile), characteristic freeze and melt rates vary between approximately $10^{-1}$ and $10^{-5}$ mm/year (Figure 3c); these are representative of a Europan setup.

This means that with a spatial map of ice-thickness observations, such as is expected from *Clipper* or *JUICE*, the spatial distribution and magnitudes of ocean freeze and melt rates should be calculable under a steady-state assumption, with a knowledge of basal viscosity. Further, the locations of maximum and minimum ocean freeze and melt rates are calculable regardless of a knowledge of basal viscosity. Finally, in the absence of a steady-state assumption, and provided that $\partial H/\partial t$ is not much larger than the other terms in Equation 8, a scaling for freeze and melt rate can still be inferred from ice-thickness observations. One advantage of using this method to infer freeze/melt rate is that it may prove useful for making inferences of ocean stratification, a control on ocean dynamics of icy satellites. This is because regions of large melt rate are regions of high rates of freshwater input; here, it may be expected that a strong two-layer ocean stratification would exist (i.e., Zhu et al. 2017).

## EARTH ANALOG RADAR VALIDATION

In order to validate our methodology, we consider radar observations of an Earth analog. Radar sounding is an active remote sensing technique that has been used extensively on both Earth and Mars to measure the thickness of ice sheets and ice shelves, leveraging the relative radio-transparency of ice (Gogineni et al. 2001; Holt et al. 2006; Vaughan et al. 2006; Plaut et al. 2007; Phillips et al. 2008). Figure 4a shows an example of data collected over Roosevelt Island, Antarctica as part of NASA's Operation IceBridge that resolves the ice surface, internal reflecting horizon, and the bedrock on which the ice rise is grounded. Given the electromagnetic wave velocity in ice, spatial and temporal variations in ice thickness can be measured directly from such data. NASA's Europa *Clipper* and ESA's *JUICE* mission will carry similar radar sounding instruments intended to study the subsurface structure and dynamics of Europa and Ganymede's ice shells, including ice shell thickness (Blankenship et al. 2017; Bruzzone et al. 2011). We show that a simple scaling of the type we propose can be used to reasonably infer accumulation rates on Roosevelt Island from ice-penetrating radar measurements of ice thickness.

### Earth-analog equations

Our icy satellite case is governed by a shear flow with a no-slip upper surface, due to the extremely cold temperatures, and a free-slip base, due to the presence of the ocean. The case of Roosevelt Island, Antarctica is similar, but with a no-slip base in contact with bedrock and a free-slip surface in contact with the atmosphere. On Earth, the relevant flow to consider is that of an ice sheet (governed by shear flow), rather than an ice shelf (which is an extensional flow, e.g., Pegler & Worster 2012; Worster 2014), even though both the Earth-based ice shelf and the ice shell of an icy satellite are floating. Here the analog to the required freezing and melting term of the ice shell needed to maintain the shell in thermodynamic balance is the accumulation of snow at the surface of the ice rise.

The equivalent governing equation for our Earth analog is then:

$$\frac{\partial H}{\partial t} - \frac{\rho_i g \gamma_e}{\eta_b}\frac{\partial}{\partial x}[H^3\frac{\partial H}{\partial x}] - b(x) = 0, \quad (13)$$

where now $\gamma_e = l[2l^2(1 - e^{-1/l}) - 2l + 1]$. Note that the difference between $\gamma_e$ in Equation 13 and $\gamma$ in Equation 8 for the icy satellite setup arises from the inverted boundary conditions. In the limit of $T_b = T_s$, $\gamma_e$ and $\gamma$ both reduce to 1/3.

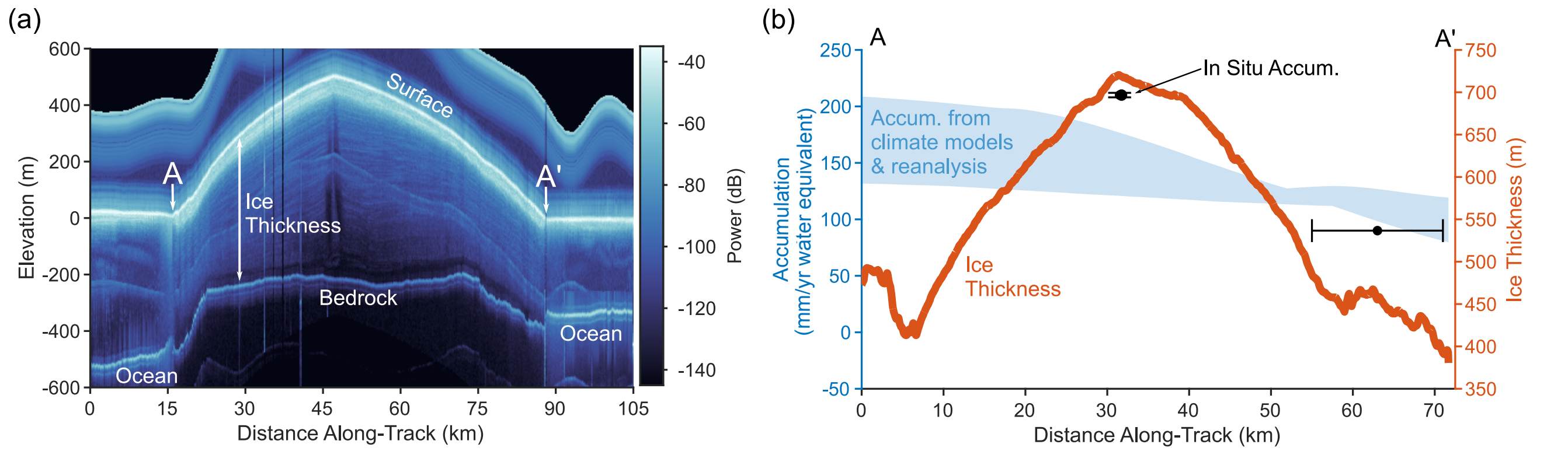

**Figure 4.** (a) Ice-penetrating radar data showing the cross-sectional structure of Roosevelt Island, a large grounded ice rise on the Ross Ice Shelf, Antarctica. (b) Relevant model values from observations. The blue shading shows the plausible range of annual accumulation rates along the transect derived from two regional climate models (RACMO, van Wessem et al. (2023) and MAR, Kittel et al. (2021); Mottram et al. (2021)) and the ERA-5 land reanalysis product (Muñoz-Sabater 2019). Sparse in-situ inferred accumulation rates are shown in the black dots (the horizontal error bars on the right dot indicate our interpretation of the uncertainty in position). Together, these products suggest that accumulation rates on Roosevelt Island range from ~80-210 mm/yr (in mm of water equivalent, which correspond to rates of about 90-230 mm/yr in ice equivalent). These are consistent with our scaling which gives estimated scalings of freeze/melt rate between 70-220 mm/year, depending on the value of $\gamma_e$. The magnitude of the annual rate of ice thickness change in mm of water equivalent ($\partial H/\partial t$) between 2003 and 2019 as measured by laser altimetry (not shown) is $\lesssim$ 31 mm/yr and typically an order of magnitude less than accumulation, suggesting that the system is in roughly steady state (i.e., $\partial H/\partial t << b$). Ice thickness, derived from the ice-penetrating radar profile in (a) assuming a dielectric constant of 3.15, is shown in the orange line.

### Validation of Scaling

On Earth, the freeze/melt rate scaling goes as:

$$b_0 \sim \frac{\rho_i g \gamma_e}{\eta_b} \frac{H_0^4}{L_0^2}. \tag{14}$$

Note the difference between this scaling and the scaling for an icy satellite, which contains a hydrostatic component related to the floating shell. Further, while freezing at the surface of Antarctica would result in meteoric ice, and freezing under an ice shell on Europa would result in congelation or marine ice (e.g., Wolfenbarger et al. 2022; Lawrence et al. 2023), it is the rate of accumulation and viscosity of ice which governs the flow dynamics, though it does seem likely that the basal viscosity, an unknown, may be related to the type of ice formed.

Estimates from Roosevelt Island radar observations suggest scalings of $H_0$ = 750 m and $L_0$ = 35 km (Figure 4a,b). We take $\rho_i$ = 920 kg m$^{-3}$, and $\gamma_e \sim 0.1 - 0.3$. (If the surface and basal temperatures vary by 10 Kelvin, for $T_b$ = 273 K, $\gamma_e \sim 0.3$, taking an activation energy $Q$ of about 60 kJ mol$^{-1}$). The viscosity of ice sheets on Earth is also not particularly well-known and much research has investigated appropriate rheology profiles for Antarctic ice sheets (e.g., Larour et al. 2005; Millstein et al. 2022; Wang et al. 2023). Here, we estimate an effective viscosity of $O(10^{14}) - O(10^{15})$ Pa s (see Materials & Methods); we assign this value to $\eta_b$. Using these parameters, we find an estimated freeze/melt scaling of ~ 70 mm/year for $\gamma_e \sim 0.1$ to ~ 220 mm/year for $\gamma_e \sim 0.3$, taking $\eta_b \sim O(10^{14})$. These rates fall in line with the estimated accumulation rates expected for Roosevelt Island (between about 100-200 mm year$^{-1}$, Bertler et al. 2018; Winstrup et al. 2019, Figure 4b). This indicates that a scaling based on gravity-driven flow dynamics and thickness/length-scale estimates may be useful for estimating accumulation rates both on Earth and on icy satellites. Similar mass-balance methodologies have also been invoked to infer melt rates at the base of Antarctic ice shelves (e.g., Wen et al. 2010; Padman et al. 2012; Adusumilli et al. 2020), further supporting this approach.

## DISCUSSION & CONCLUSION

### Summary

On long time scales, ice flows as a viscous fluid (e.g., Pegler & Worster 2012; Worster 2014). Here we describe the dynamics of gravity-driven ice shell flow and describe how this can be used to infer melt and freeze rates under the ice shells of icy satellites, which are currently impossible to measure directly. Since the ice shell of an icy satellite experiences a steep temperature drop from its base to its surface, and viscosity depends exponentially on temperature, we formulate the viscous gravity current equations to include a temperature-dependent viscosity. This shows analytically how the viscous flow of the ice shell is confined to its base; the thickness of the viscous portion of the shell depends on the across-shell temperature difference. In the limit of no temperature jump, our equation reduces to the classical gravity current equation with no-slip and free-slip boundary conditions (Huppert 1982).

We describe how a balance between ice flow and freeze and melt rate implies a scaling for freeze and melt rate which depends on a vertical ice thickness scale and a horizontal length scale. We further describe how in steady state, ocean freeze and melt rates and the spatial locations of maximum freeze and melt rates can be calculated from ice-thickness measurements. Inferences from radar observations from Roosevelt Island, Antarctica are used to corroborate our scaling and give credence to the methodology we propose. We expect that with this methodology, ocean parameters such as freeze and melt rates may be inferred with future measurements of ice thickness from upcoming space missions.

**Are Europa and Enceladus in Steady State?**

Much past work, whether considering the dynamics or thermodynamics (or both) of Europan and Enceladean ice shells, assumes that the ice shells have reached steady state (e.g., Ojakangas & Stevenson 1989; Hussmann et al. 2002; Tobie et al. 2003; Ashkenazy et al. 2018; Hemingway & Mittal 2019; Akiba et al. 2022; Kang & Jansen 2022; Kang 2022). Nonetheless, it is not actually clear if these shells can be considered to be in equilibrium (see Shibley & Goodman 2024, for a thermodynamic case on Europa) and (Čadek et al. 2019, for a dynamic case on Enceladus).

If the steady-state assumption does not apply, then it would not be possible to directly calculate a spatially-varying freeze and melt rate as described. However, the freeze and melt rate scaling will still be applicable provided that the term $\partial H/\partial t$ is not much larger than the nonlinear diffusive term or the freeze and melt term in Equation 8 (See Section B of Materials & Methods for more discussion). Such a system is exemplified via the radar validation of Roosevelt Island, Antarctica in the previous section which is not strictly in steady state, but for which the scaling holds. We note that at timescales appropriate for the surface age of Europa, $\partial H/\partial t$ is a similar order of magnitude to the nonlinear diffusive term for lower values of viscosity (i.e., $\sim 10^{14}$ Pa s) and larger thickness scales $H_0$; higher $\eta_b$ and lower $H_0$ increase the timescale of flattening for the system (see Section B of Materials & Methods).

**What does Freeze and Melt Rate Tell You?**

Ultimately, an understanding of ocean freeze and melt may give insight into ocean stratification. Stratification is a key control on an expected ocean circulation and thus the transport of nutrients and other tracers that may be of astrobiological interest (see Lobo et al. 2021, who describe this for the case of Enceladus). Consider for example a region of high melt rate. This would result in a large influx of freshwater into a particular region of the ocean, likely generating a strong stratification and depressing isopycnals (surfaces of constant density). A similar description of a freshwater-stratified region and the circulation it implies via conservation of salt and heat has been described in Zhu et al. (2017). Our methodology provides a framework to infer freeze and melt rates, which are otherwise challenging to measure under kilometers of ice, from observations of ice thickness; these rates ultimately relate to stratification.

An important point for future work is to help constrain the basal viscosity of the ice shell. This has been a limitation of modeling studies of ice shells of icy satellites (including our own), as the rheology controls the relevant dynamics but is not well-defined (e.g., Shibley & Goodman 2024). A particular issue is that the elastic processes on the surface can not be easily inverted via a temperature-dependence to constrain the viscous processes of the base. Developing a theoretical or observational way to constrain the basal viscosity is a key area for future research.

## MATERIALS AND METHODS

### A. Numerical Method

Equation (8) is a one-dimensional partial differential equation that is solved numerically subject to the following conditions:

$$\frac{\partial H}{\partial x}|_{x=L} = 0, \text{ no flux at the equator, and} \tag{15}$$

$$\int_{x=0}^{x=L} H(x,t)dx = V, \text{ volume conservation} \tag{16}$$

where $L$ is the end of the horizontal domain. We use a forward difference scheme in time and a centered-difference scheme in space. The spatial grid step is 10 km, and the time step is $5 \times 10^8$ seconds. We set up a grid of 241 points, equivalent to a horizontal length of $2.4{\times}10^6$ m, approximately the distance from Europa's pole to equator.

We approximate the initial thickness of the ice shell as a complementary error function profile:

$$H(x,0) = h_0 + a\frac{2}{\sqrt{\pi}}\int_{cx-2}^{\infty} e^{-y^2}dy, \tag{17}$$

where $h_0$ is some initial background thickness, $a = h_0/8$, and $c$ is $2.8{\times}10^{-6}$ m$^{-1}$. The constants $h_0$, $a$, and $c$ can be changed to approximate different initial thickness profiles. In the figures presented here (aside from Figure 3c, where $h_0$ varies), we take $h_0 = 8$ km. This functional form is chosen to simulate an ice shell which was thicker at the pole and thinner at the equator, whose thickness decreased between pole to equator in some reasonable way. An initial pole thickness of about 10 km is taken since this is expected to be a plausible estimate for a conductive ice shell on Europa.

### B. Flattening Timescale

The timescale at which an ice sheet flattens depends on the viscosity and on the thickness scale of the ice shell. This follows the scaling law:

$$T_0 \sim \frac{\eta_b}{\gamma \rho_i g}\left(\frac{\rho_w}{\rho_w - \rho_i}\right)\frac{L_0^2}{H_0^3}, \tag{18}$$

where $T_0$ is a characteristic time scale, $H_0$ is a characteristic thickness, $L_0$ is a characteristic horizontal length scale, $\rho_w$ is the density of ocean water, $\rho_i$ is the density of ice, $g$ is gravity, and $\eta_b$ is the basal ice viscosity. In order for the steady-state assumption to apply here requires $T_0 >> H_0/b_0$ (or equivalently $T_0 >> \frac{\eta_b}{\rho_i g \gamma}(\frac{\rho_w}{\rho_w-\rho_i})\frac{L_0^2}{H_0^3}$). Thus we can find a scaling for both the time at which a system subject to freezing and melting can be approximated by a steady-state, and a scaling for freeze and melt rate which would maintain the ice thickness gradient.

Our assumption of a linear temperature profile implies that possible internal shell heating does not significantly affect the temperature profile; we expect that the effect of including internal heating in our conductive setup would essentially reduce a depth-integrated viscosity in the lower portion of the shell (by keeping temperatures higher), likely leading to somewhat faster ice flow and thus increased freeze/melt rates.

### C. Estimate of Viscosity

Following the established convention in terrestrial glaciology (Glen 1958; Nye 1957), we define an effective viscosity as:

$$\eta = \frac{B}{2}\dot{\epsilon}^{(1/3-1)}, \tag{19}$$

where $\dot{\epsilon}$ is the strain rate, and $B$ is defined as $B(T) = 2.207\exp(\frac{3155}{T} - \frac{0.16612}{(273.39-T)^{1.17}})$ (Hooke 1981; van der Veen 1998), where $T$ is temperature.

Estimates for $B(T)$ for two extreme temperatures are as follows:

(1) The warmest ice is near melting temperature at around $T = 270$ K (Thomas et al. 1980), yielding $B(T) \sim 250,000$ Pa year$^{1/3}$. (2) The surface temperature in this area is about $T = 247$ K (Thomas et al. 1980), yielding $B(T) \sim 800,000$ Pa year$^{1/3}$.

Finally, the order of magnitude of the effective strain rate $\dot{\epsilon}$ on Roosevelt Island as derived from satellite measurements of surface ice velocities is about $5\times10^{-4}$ year$^{-1}$. Based on the upper and lower bounds of $B$, the effective viscosity can vary in the range of $\eta = 6 \times 10^{14} - 2 \times 10^{15}$ Pa s.

## ACKNOWLEDGEMENTS

N.C.S., C.-Y.L., and R.C. acknowledge internal funding support from Princeton University. N.C.S. acknowledges a postdoctoral fellowship from the Princeton Center for Theoretical Science. R.C. acknowledges the Harry Hess Postdoctoral Fellowship from the Princeton Department of Geosciences. C.-Y.L. acknowledges internal funding from Stanford University, and R.C. acknowledges support from Cornell University. N.C.S. acknowledges helpful conversations with Jeremy Goodman and Glenn Flierl. This work was performed in part at the Aspen Center for Physics, which is supported by National Science Foundation grant PHY-2210452. This work was partially supported by a grant from the Simons Foundation.

## DATA AVAILABILITY

The numerical method to solve Equation 8 is given in the Materials and Methods appendix of the text.

The ice-penetrating radar data shown in Figure 4 are available from the University of Kansas Center for Remote Sensing and Integrated Systems (last access: 2023-12-08, Leuschen et al. (2023)). (We acknowledge the use of data and/or data products from CReSIS generated with support from the University of Kansas, NASA Operation IceBridge grant NNX16AH54G, NSF grants ACI-1443054, OPP-1739003, and IIS-1838230, Lilly Endowment Incorporated, and Indiana METACyt Initiative.)

The rate of ice-thickness change ($\partial H/\partial t$) can be inferred from the data at University of Washington Research Works (last access: 2023-12-08) (Smith et al. 2020).

The accumulation range in Figure 4b was generated in part using ERA5-Land reanalysis products, which are available from the Copernicus Climate Data Store (Muñoz-Sabater 2019, last access: 2023-12-08). This contains modified Copernicus Climate Change Service information 2023. Neither the European Commission nor ECMWF is responsible for any use that may be made of the Copernicus information or data it contains.

MAR regional climate simulations are available from Zenodo (last access: 2023-12-08) (Kittel et al. 2021; Mottram et al. 2021). We thank the MAR team which make available the model outputs, as well agencies (F.R.S - FNRS, CÉCI, and the Walloon Region) that provided computational resources for MAR simulations.

RACMO regional climate simulations are available from Zenodo (last access: 2023-12-8) (van Wessem et al. 2018; van Wessem et al. 2023). In situ inferred accumulation rates come from Bertler et al. (2018) and Winstrup et al. (2019).